\def\year{2023}\relax
\documentclass[letterpaper]{article} 
\usepackage{aaai22}  
\usepackage{times}  
\usepackage{helvet}  
\usepackage{courier}  
\usepackage[hyphens]{url}  
\usepackage{graphicx} 
\usepackage{booktabs}
\usepackage{natbib}  
\usepackage{caption} 
\DeclareCaptionStyle{ruled}{labelfont=normalfont,labelsep=colon,strut=off} 
\usepackage{algorithm}
\usepackage{algorithmic}
\usepackage{array,multirow}
\usepackage{newfloat}
\usepackage{listings}
\floatstyle{ruled}
\newfloat{listing}{tb}{lst}{}
\floatname{listing}{Listing}
\usepackage{color}

\title{Characteristics of Coordinated Inauthentic Accounts}
\title{Socio-Linguistic Characteristics of Coordinated Inauthentic Accounts}
\author{Anonymous Author(s)}
\author {
    Keith Burghardt,\textsuperscript{\rm 1}
    Ashwin Rao, \textsuperscript{\rm 2}\equalcontrib
Siyi Guo, \textsuperscript{\rm 2}\equalcontrib
Zihao He, \textsuperscript{\rm 2}\equalcontrib
Georgios Chochlakis, \textsuperscript{\rm 2}\equalcontrib
Baruah Sabyasachee, \textsuperscript{\rm 2}\equalcontrib
Andrew Rojecki, \textsuperscript{\rm 3}
Shri Narayanan, \textsuperscript{\rm 2}
Kristina Lerman \textsuperscript{\rm 1}    
}
\affiliations {
    \textsuperscript{\rm 1} USC Information Sciences Institute\\
    \textsuperscript{\rm 2} University of Southern California\\
    \textsuperscript{\rm 3} University of Illinois Chicago\\
   \{keithab,ashreyas,siyiguo,zihaohe,chochlak\}@isi.edu, sbaruah@usc.edu, arojecki@uic.edu, \{shri,lerman\}@isi.edu
   }

\begin{document}

\maketitle             

\begin{abstract}
Online manipulation is a pressing concern for democracies, but the actions and strategies of coordinated inauthentic accounts, which have been used to interfere in elections, are not well understood. We analyze a five million-tweet multilingual dataset related to the 2017 French presidential election, when a major information campaign led by Russia called ``\#MacronLeaks'' took place. We utilize heuristics to identify coordinated inauthentic accounts and detect attitudes, concerns and emotions within their tweets, collectively known as \emph{socio-linguistic characteristics}. We find that coordinated accounts retweet other coordinated accounts far more than expected by chance, while being exceptionally active just before the second round of voting. Concurrently, socio-linguistic characteristics reveal that coordinated accounts share tweets promoting a candidate at three times the rate of non-coordinated accounts. Coordinated account tactics also varied in time to reflect news events and rounds of voting. Our analysis highlights the utility of socio-linguistic characteristics to inform researchers about tactics of coordinated accounts and how these may feed into online social manipulation.
\end{abstract}

\section{Introduction}

Social media platforms are potent vectors for manipulation~\cite{bradshaw2018global,badawy2018analyzing,kim2018uncover}. Malicious actors use Facebook, Twitter, and other platforms to deploy inauthentic accounts that interact with and manipulate authentic users on each side of a divisive issue~\cite{ratkiewicz2011detecting,kim2018uncover}, recruit converts, incite violence, spread misinformation~\cite{vosoughi2018spread}, or undermine trust in democratic institutions~\cite{badawy2019characterizing}. Although these platforms have invested heavily to remove harmful accounts, malicious actors have adapted their strategies to evade detection and develop increasingly sophisticated influence campaigns \cite{Sayyad2020}. Technologies have been developed to detect, characterize, and track inauthentic account activity at scale \cite{Ferrara_2017,Sayyad2020,omsome2022_ukraine}, but there is a pressing need to better understand the tactics and strategies of influence campaigns that utilize inauthentic accounts through analysis of the content they promote.

In this paper, we analyze a large corpus of over 5M tweets related to the 2017 French presidential election to identify influence campaigns intended to affect the outcome of the election~\cite{Ferrara_2017}. We use a state-of-the-art heuristic to identify \textit{coordinated inauthentic accounts} \cite{Pacheco2021} (we call these ``coordinated accounts'' for brevity) that may be attempting to influence election outcomes. 
We then create computational methods to identify attitudes, concerns, and emotions within influence campaigns. We define \textit{attitudes} as the opinion of a user, \textit{concerns} as the issues discussed, and \textit{emotions} as the feelings expressed in text. Finally, we analyse how the coordinated accounts utilize these features to inform us about their tactics. 

We study the French presidential election cycle, which kicked off on 10 April 2017. The first round of voting took place on 22 April 2017, with Emanuel Macron and Marine Le Pen advancing to the second round. Our motivation to analyze the 2017 election in particular is because there was a leak of French presidential candidate Emmanuel Macron's campaign emails (\#MacronLeaks) on 5 May, just before the second round of voting on 7 May. \#MacronLeaks leveraged a large cache of hacked documents and emails shared on WikiLeaks to discredit Macron and his party, En Marche~\cite{Ferrara_2017,vilmer2021fighting}, likely orchestrated by Russia \cite{Gray2022}. It was exposed on the imageboard 4Chan and tweeted on 5 May by American alt-right activist Jack Posobiec \cite{Gray2022}. Although the campaign ultimately failed to achieve its presumed goal (as Macron won the second round of voting) the campaign acts as an important case study of coordinated account tactics. The coordinated accounts we find are strongly over-represented in the \#MacronLeaks tweets, as they were only 0.28\% of all accounts but represented at least 18.7\% of tweets with hashtags related to the leak within our dataset, which could represent an attempt to influence the election.

We next hypothesize a range of tactics coordinated accounts utilize through analysis of socio-linguistic characteristics. The unusual prevalence (or lack) of particular socio-linguistic characteristics within coordinated accounts compared to non-coordinated accounts helps us understand what coordinated accounts attempt to promote. The differences in between clusters of coordinated accounts, meanwhile, help us distinguish unique tactics that some clusters of coordinated accounts use that others do not. For example, one cluster of coordinated accounts heavily promoted concerns about national pride, international alliances, while another appeared to discuss the president of Gabon with no mention of French campaign issues. This is suggestive of multiple competing influence campaigns happening during the French election. We then show how the frequency of socio-linguistic characteristics changes over time to identify tactics, such as promoting candidates just before an election.
Finally, we show how the prevalence of particular languages in each cluster hint at the different audiences for each influence campaign, such as the use of English within the pro-Marine Le Pen cluster of coordinated accounts versus French within the pro-Benoît Hamon and Francis Fillon clusters, who were round one presidential candidates. Twitter is used in France in much the same way it is used many areas of the world (e.g., for social interactions, news, political discourse, etc.), even in elections  \cite{nooralahzadeh20132012}, thus we believe our results will generalize well outside of this election scenario.

To summarize, our contributions are the following:
\begin{itemize}
    \item We develop novel multilingual techniques to detect socio-linguistic characteristics from tweets and make our entire pipeline publicly available.
    \item We use three techniques to extract coordinated networks of inauthentic accounts from Twitter users in a major election, and publicly share this code. 
    \item We extract coordinated account behaviors and socio-linguistic characteristics.
    \item We apply our findings to hypothesize influence tactics.
\end{itemize}
Overall, our analysis demonstrates the feasibility of automatically identifying potential tactics used in online influence campaigns. Our code, human annotations, and example coordinated tweets are shown in the following repository: \url{https://github.com/KeithBurghardt/Coordination/}.

\section{Related Work}

\paragraph{Political Manipulation} 

Online manipulation  is a worldwide phenomenon (cf. \cite{tucker2018social} for a review), and can occur through a variety of ways, such as search ranking or social media trend manipulation. We specifically focus on inauthentically sharing posts that have a particular frame, a prototypical example of online manipulation \cite{tucker2018social}. This type of manipulation has long been explored on social media~\cite{ratkiewicz2011detecting,ratkiewicz2011truthy,kim2018uncover}, including the Brexit vote~\cite{howard2016bots}, the 2016 US presidential election~\cite{bessi2016,badawy2018analyzing,badawy2019characterizing,kim2018uncover}, the 2017 French elections~\cite{Ferrara_2017}, and the 2022-2023 Russia-Ukraine war \cite{omsome2022_ukraine}.  The impact of these accounts is uncertain \cite{Freelon2020}, but engagement with, and attempts to manipulate, authentic users is of grave concern. 

\paragraph{Coordinated Inauthentic Accounts}
Several studies also focus on the detection and behavior of coordinated accounts in social media, including on Facebook \cite{Giglietto2020,Giglietto2020Facebook}, YouTube \cite{kirdemir2022towards}, and Twitter \cite{Sharma2020,Nizzoli2021,Weber2021,Mazza2022,Cinelli2022}. In contrast to bot or troll detection, coordinated account analysis focuses on detecting and analyzing accounts working in concert \cite{starbird2019disinformation}. Ways to uncover coordinated accounts include temporal similarities in users \cite{Sharma2020,Weber2021,Schliebs2021,Pacheco2021}, similarity in content \cite{Schliebs2021}, comment networks \cite{kirdemir2022towards}, URLs shared \cite{Giglietto2020Facebook}, user attributes, and co-retweeting \cite{Pacheco2021,Mazza2022}.

Most of these studies analyze these coordinated campaigns within elections, although there are exceptions to this trend, such as coordinated accounts related to COVID-19 \cite{Graham2020,PinaGarcia2022}. The goals of coordinated accounts, however, are less-studied. While previous work includes analyzing stories promoted by coordinated accounts \cite{Ehrett2021}, or stances by social bots \cite{CHEN2021913}, there is a lack of research on socio-linguistic characteristics expressed by coordinated accounts, and how they may feed into manipulation tactics.

\paragraph{Attitude Analysis}
Attitudes, such as voting for or against a candidate are a distinct set of tools we utilize in this paper, but have analogues in previous work. Attitudes most closely resemble stances (for a review, cf. \cite{kuccuk2020stance}), previously used to study misinformation \cite{hardalov-etal-2022-survey}, as they aim to determine the opinions users are trying to convey. Meanwhile, some attitudes, such as the belief that a candidate is corrupt, are similar to moral framing \cite{Linvill2021}, whereby an action is viewed as a virtue or vice, or person is viewed as virtuous or corrupt. 

\paragraph{Concern Analysis}
Concerns, meanwhile, represent key topics discussed by Twitter users, and have analogues to topic modeling \cite{Mei2007,eisenstein2011sparse,Jelodar2019}, framing \cite{card2015media}, as well as position issues~\cite{stokes1963} that divide voters. Among the many possible topics, we focus on those discussed by the French presidential election~\cite{lachat2020campaigning}.

\paragraph{Emotion Analysis}
Emotion extraction tools have perhaps the longest history, starting with the General Inquirer \cite{stone1966general}, and were iteratively improved with dictionary-based methods, such as LIWC \cite{pennebaker2001linguistic}, 
EmoLex \cite{mohammad2010emotions}, and  DDR \cite{garten2018dictionaries}.
Alike to dictionary-based methods, bag-of-words features have been used along side other features to build emotion recognition systems \cite{wang2015detecting,Li2015}, including sentence-level emotion predictions \cite{li2015sentence}.

The most successful emotion recognition methods deploy Deep Learning \cite{he2018joint}, such as those based on LSTMs \cite{hochreiter1997long,duppada2018seernet,yu2018improving}, and bidirectional LSTMs \cite{baziotis2018ntua}. 
Recently, Transformers \cite{vaswani2017attention} have dominated the field. Ying et al., (\citeyear{ying2019improving}) use the \verb+[CLS]+ token of BERT along with a shallower Convolutional Neural Network meant to learn task-specific n-gram patterns to predict emotions. Other methods use Graph Neural Networks \cite{xu2020emograph}, and took advantage of correlations between emotion \cite{alhuzali-ananiadou-2021-spanemo}. In this paper, we use the current state-of-the-art algorithm, Demux \cite{Chochlakis2023}, which outperforms competing methods on the SemEval 2018 Task 1 e-c benchmark \cite{SemEval2018Task1}. 

\section{Methods}
We present the data and the methods we use to extract socio-linguistic characteristics from tweets.  

\subsection{Data}
We apply our methods to a corpus of 5.3M tweets about the 2017 French presidential election and automatically detected the attitudes, concerns, and emotions in each tweet. 
The tweets were collected by querying Twitter with a set of keywords related to the election: e.g., ``election'', ``élection'', ``l'élection'', ``Elysee 2017'', ``Elysee2017'', etc.~\cite{Ferrara_2017}. In addition, collected tweets include those posted by accounts of presidential candidates, their parties or campaigns, such as @MLP\_officiel, @EmmanuelMacron, @enmarchefr, @JLMelenchon, and @jlm\_2017. The vast majority (91\%) of tweets were in French, 4\% were in English, and the rest were a wide variety of other languages including 3\% unknown based on the Twitter API's language detection feature. 

\begin{figure}
    \centering
    \includegraphics[width=\columnwidth]{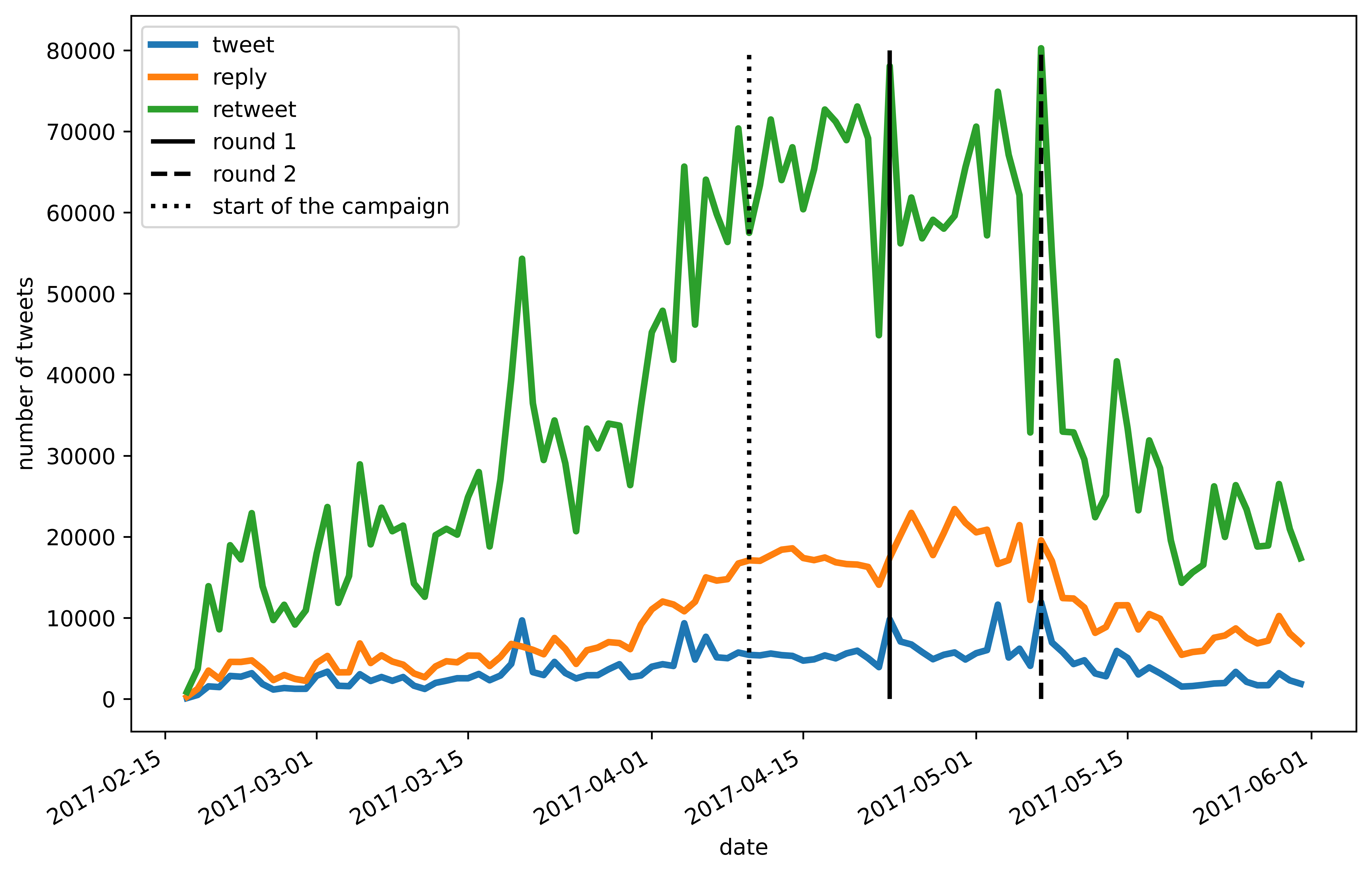}
    \caption{Daily number of tweets, retweets and replies in the 2017 French election data. Vertical lines mark important events: start of the campaign (dotted line), first round of voting (solid line), second round of voting (dashed line).}
    \label{fig:TweetsOverTime}
\end{figure}

Fig.~\ref{fig:TweetsOverTime} shows the daily volume of messages. Online discussion geared up long before the official start of the presidential campaign (10 April 2017) with sharp peaks on the days of the first (23 April 2017) and second (7 May 2017) rounds of voting. Interest in the campaign dropped sharply thereafter, with small increases around the time of Macron's inauguration (14 May 2017). Although quote tweets were used in 2017, they are missing in our data, therefore in the rest of the paper, we analyze original tweets, replies, and retweets.

\subsection{Attitude Detection} 
Attitudes describe what a message's author thinks and believes. In the context of an election, influence messages express attitudes that promote a candidate or party either by explicitly telling voters to vote for or against them or by using moral outrage (e.g., saying a candidate is immoral) to drive people from opposing candidates and parties. Moral framing, such as framing a candidate or party as corrupt \cite{Linvill2021}, is a powerful motivator strongly linked to partisan identity~\cite{graham2018moral}. 

\noindent \paragraph{Vote for or Against CANDIDATE or PARTY}

To detect the author's attitude towards the target, i.e., CANDIDATE or PARTY,  we frame the problem as stance detection (cf. \cite{kuccuk2020stance,hardalov-etal-2022-survey})
. The detected stance can be ``in favor,''  ``against'' or neutral: e.g., if we find that a tweet is in favor of Macron, then its attitude is ``vote for Macron''. Here, CANDIDATE or PARTY is a wildcard that represents any of the 11 candidates in 2017 presidential election or their associated parties, including the run-off candidates Macron (party: En Marche) and Le Pen (party: Front National, later renamed Rassemblement National in 2018).

We encode each pair consisting of a text (tweet) and a target (CANDIDATE/PARTY) with a pretrained multilingual text embedding model, XLM-T \cite{xlm2021barbieri}. This representation is then fed into a feed-forward neural network for stance classification. This intuitive method poses two challenges. First, inferring the stance entails some background knowledge about the target; second, tweets labeled with stances towards candidate in the 2017 French Election are scarce, making  supervised learning difficult.

To address the first challenge, we use a stance detection model WS-BERT~\cite{he2022infusing} that uses relevant Wikipedia entries for background information about the target needed to infer stance. By using XLM-T embeddings, instead of BERT used previously~\cite{he2022infusing}, however, our method can extend to multilingual data. To meet the second challenge, we pre-train the model on two other supervised stance detection datasets, COVID-19-Stance \cite{glandt-etal-2021-stance} and P-Stance \cite{li2021improving}, and then fine-tune this model on 10K human annotated tweets, described later. We apply this model to infer the stance about 11 candidates. COVID-19-Stance has tweets in a COVID-19 domain annotated with ``favor'' and ``against'' for ``Anthony S. Fauci, M.D.'', ``Keeping Schools Closed'', ``Stay at Home Orders'', and ``Wearing a Face Mask.'' P-Stance has tweets in a political domain annotated with ``favor'' and ``against'' for ``Biden'', ``Sanders'' and ``Trump'', which lies in a political domain similar to our case.




\noindent \paragraph{CANDIDATE or PARTY is Moral or Immoral}

We operationalize moral judgment using Moral Foundations (MF) Theory \cite{graham2013moral}, which proposes five dimensions of morality, each with its virtues and vices: care vs. harm, fairness vs. cheating, loyalty vs. betrayal, authority vs. subversion, and sanctity vs. degradation. We consider all the virtues to define the class ``moral,'' and all the vices as the class ``immoral.''

For this model, we first pre-process all tweets by removing URLs, replacing all mentions with ``@user'', removing or split hashtags, converting emojis to a  description, converting text to lower case removing punctuations and non-ascii text, and removing emoticons.

We then train our model first using Moral Foundations Tweet Corpus (MFTC)~\cite{hoover2019mftc}, which contains English language tweets annotated by the morality they express, and then fine-tune the model with 10K human annotated French tweets. For each tweet, we take majority vote as the true label. We then fine-tune a pre-trained multilingual model XLM-T \cite{xlm2021barbieri} with a binary prediction layer (a sigmoid activation). The model allows for multi-label prediction, because a tweet may express more than one moral judgment. We further finetune this model using 10K human annotated tweets. Although we do not have an equivalent French moral dictionary, the XLM-T multilingual embedding allows our model to transfer knowledge from English words to the majority-French dataset. 

\subsection{Concern Detection}
Concerns are divisive issues that separate potential voters into distinct blocs, i.e., position issues~\cite{stokes1963}. We focus on a subset of the issues salient to the 2017 French presidential election~\cite{lachat2020campaigning}, namely Economy, Terrorism, Religion, Immigration, International Alliances. Russia Relations, National Identity, Environment, Misinformation, and Democracy.



To detect concerns, we fine-tune a BerTweetFr model \cite{guo-etal-2021-bertweetfr} to predict concerns from 10K human annotated data using the AutoModelForSequenceClassification option from HuggingFace \cite{auto} and train for 3 epochs with a batch size of 8. Each concern becomes a binary label prediction task, allowing for multiple concerns to be found in each tweet. 


\subsection{Emotion Detection}
Emotions are feelings expressed in a message. Even a short text---a tweet---can convey emotions. The emotional expression spans a range from anger and hate to joy and pride.

Our emotion detection tool is based on Demux \cite{Chochlakis2023}, which is the state-of-the-art model trained on SemEval 2018 Task 1 E-c (extracting emotions from text) \cite{SemEval2018Task1}. Demux includes the names of emotions in the input as its first input sequence, and the actual input as the second sequence. The contextual embeddings for each emotion are used to get a confidence. 
Consequently, the model can predict none, one, or multiple emotions per input. We apply XLM-T \cite{xlm2021barbieri,huggingxlmt} to Demux to improve multi-lingual emotion prediction.

To simplify emotion recognition, similar emotions that often co-occur are grouped into clusters. Our approach attempts to automatically recognize these clusters: ``Anger, Hate, Contempt and Disgust'', ``Embarrassment, Guilt, Shame and Sadness'', ``Admiration and Love'', ``Optimism and Hope'', ``Joy and Happiness'', ``Pride and National Pride'', ``Fear and Pessimism'', ``Amusement'', other positive emotions, and other negative emotions. These labels combine similar emotions, and account for nuances of the French election (e.g., discussion of pride, including national pride). Amusement meanwhile is not an emotion per se, but we find is often evoked in tweets. 

Using the English and Spanish tweets in SemEval 2018 Task 1 E-c for pre-training \cite{duppada2018seernet}, we combined anger and disgust into ``Anger, Hate, Contempt and Disgust''; sadness into ``Embarrassment, Guilt, Shame and Sadness''; love into ``Admiration and Love''; optimism into ``Optimism and Hope''; joy into ``Joy and Happiness''; and fear and pessimism into ``Fear and Pessimism''.  The other labels were not pre-trained. Due to the multilingual nature of these embeddings, pre-training on non-French data does not harm the model. We then fine-tuned the model with 10K human annotations of French tweets, which have support over all the emotions. 

\subsection{Fine-Tuning And Evaluation Dataset}

An independent Testing \& Evaluation (T\&E) team is used to annotate 10K  French election tweets. 
The T\&E team recruited and trained 15  annotators who were all fluent French speakers and actively followed French politics. They were given an annotation guide document written in English (shown in \url{https://github.com/KeithBurghardt/Coordination/tree/main/annotations}), describing what each attitude (called an ``agenda'' in the document), concern, and emotion represents. Annotators were given a small subset of these 10K tweets such that at least three annotators labeled each tweet for each socio-linguistic characteristic. These labels were all binary and a tweet could contain multiple attitudes, concerns, or emotions. 
The unweighted mean inter-annotator agreement, $\kappa$ \cite{Cohenkappa} is 0.51 for attitudes, 0.67 for concerns, and 0.34 for emotions, which represents fair to substantial agreement \cite{kappa}.

\begin{figure}[thb!]
    \centering
    \includegraphics[width=\columnwidth]{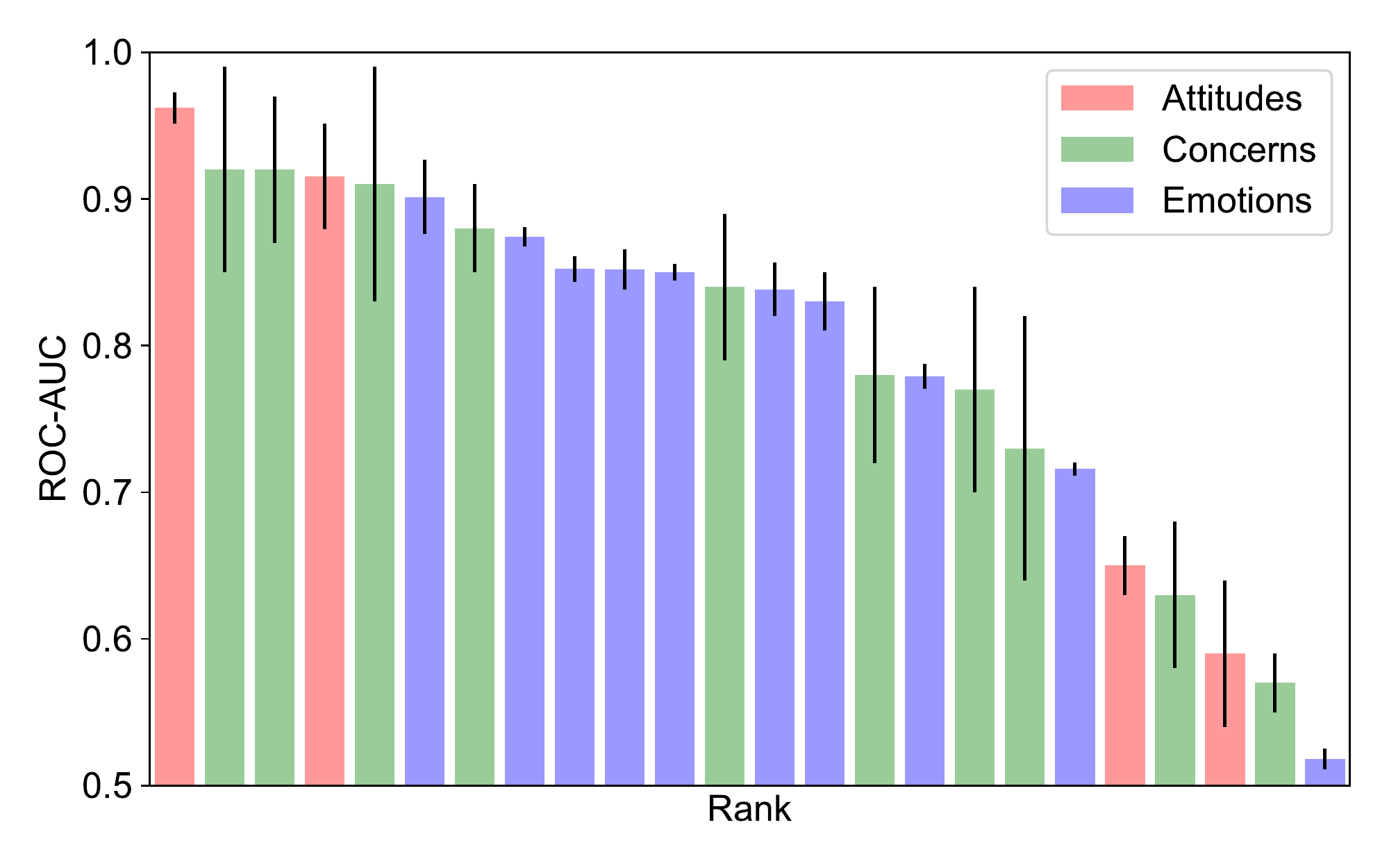}
    \caption{Evaluation of the models' predictions on a 5K subset of 2017 French Election tweets. The bars show AUC scores predicted by the models, ranked from highest to lowest ROC-AUC in held-out data: Vote for attitude, Economy and Terrorism/counterterrorism concerns, Vote against attitude, Religion concern, Embarrassment emotion, Immigration concern, Anger, Pride, Sarcasm, and Hope emotions, Environment concern, Admiration and Fear emotions, Misinformation concern, Positive-other emotion, Russia and Democracy concerns, Negative-other emotion, Immoral attitude, National identity concern, Moral attitude, International alliances concern, and the Joy emotion. Black lines indicate standard errors after bootstrapping ten times (see Methods). All results are statistically significantly above the 0.5 baseline (z-score p-value $<0.05$) except for Morals (p-value$=0.07$).}
    \label{fig:indicators_AUC}
\end{figure}

To evaluate the models, we reshuffle these 10K tweets and take the first 5K for training while holding out the next 5K for testing. We then compute the ROC-AUC for each attitude, concern, and emotion. This process is repeated ten times to calculate the variance of the performance metrics. The results are shown in Fig.~\ref{fig:indicators_AUC}. Our models generally achieve high ROC-AUC scores, which gives us confidence in their ability to detect these features.

\subsection{Coordinated Inauthentic Accounts}
 Coordinated accounts are accounts that work together towards some broader objective while seeking to mislead people about their goals~\cite{Giglietto2020,Pacheco2021,Cinelli2022}. Such accounts could be social bots~\cite{ferrara2016rise}, or humans, e.g., paid trolls~\cite{badawy2018analyzing}. 
 Due to the Twitter terms of service that we follow, we can not check if accounts are bots as all data, including usernames, are anonymized. Moreover, even if the data were not anonymized, the high false positive rate for bot detection \cite{Rauchfleisch2020} makes insights about bots more difficult to infer. To collect networks of coordinated accounts, we identify pairs of accounts with unexpectedly similar behaviors~\cite{Nizzoli2021}, namely those whose original tweets share five or more hashtags in the same order, which represents tweets that are semantically very similar. We do not claim that this method creates an exhaustive list of coordinated accounts in the dataset. However,  this heuristic can detect the largest number of likely coordinated accounts compared to alternative methods~\cite{Pacheco2021}, such as timing of messages, sharing  user profile information, sharing of what is retweeted, and other features~\cite{Giglietto2020}. For robustness, however, we compare this method against two alternatives: retweet similarity and tweet time similarity. The former is defined as taking a TF-IDF vector of all tweets that are retweeted in the dataset. The top 0.5\% of cosine similar users that have more than ten retweets in the dataset are considered coordinated. We will show that this method has drawbacks. We contrast this method with tweet time similarity. To calculate this metric, we first extract the time any tweet (original, reply, or retweet) was sent for each account that has sent more than ten tweets. We bin these tweets into 30 minute intervals, and convert the series of binned tweet times for each account into a TF-IDF vector. If the cosine similarity of these accounts is $>0.99$ (this is an arbitrary cutoff; results are robust to this choice) then we consider the accounts coordinated.

\section{Results}

We extract socio-linguistic characteristics of tweets to study user behavior during the election cycle, how people respond to external events, and to elucidate coordinated account tactics within information campaigns. 

\begin{figure}[tbh!]
    \centering
    \includegraphics[width=\columnwidth]{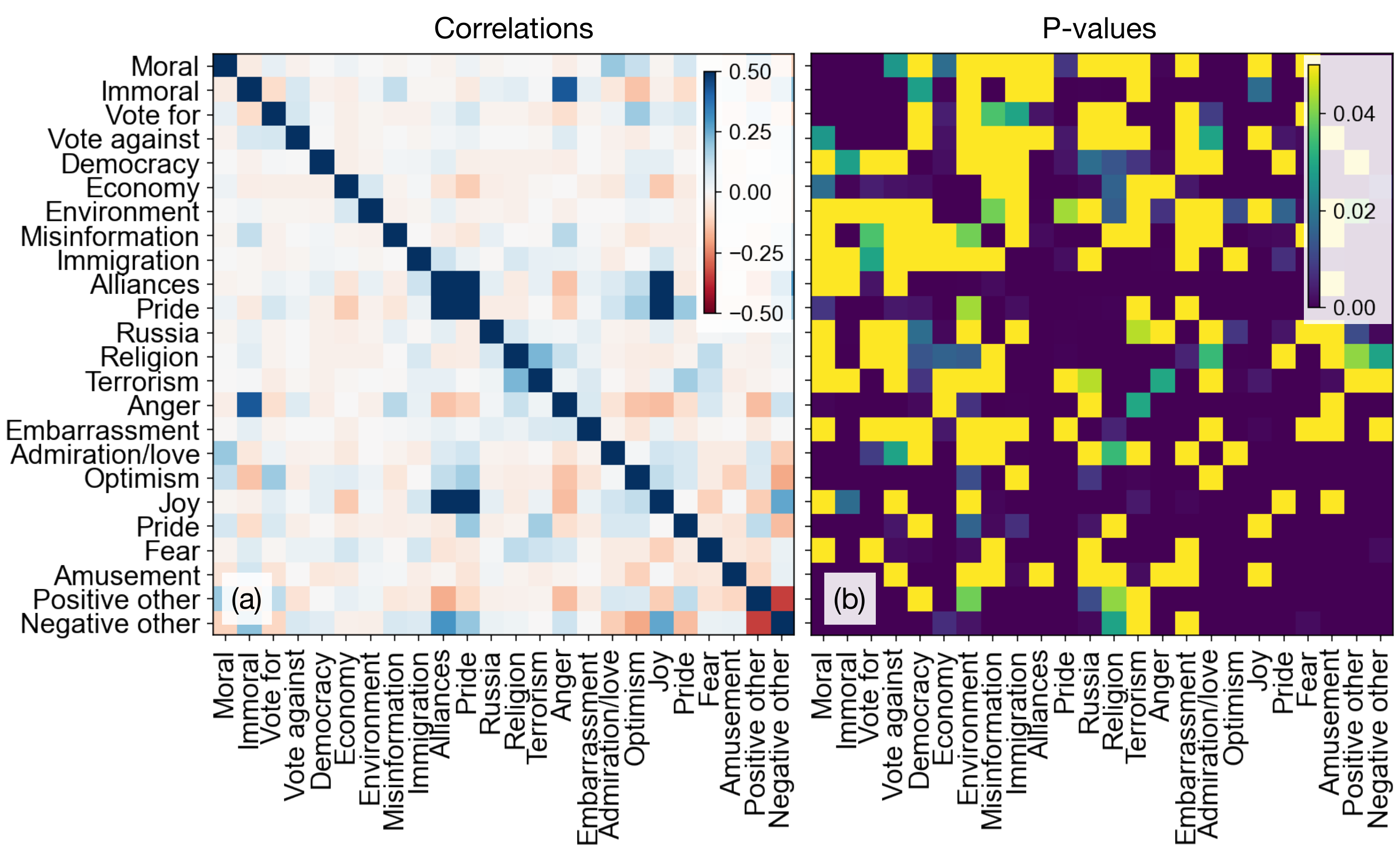}
    \caption{Spearman rank correlations between attitudes, concerns, and emotions for 10K human annotated tweets. (a) Correlations and (b) p-values of the correlations.}
    \label{fig:correl}
\end{figure}

\subsection{Correlation of Socio-Linguistic Characteristics}
We first spot check the validity of the socio-linguistic characteristics by analyzing their correlations. Figure~\ref{fig:correl}a shows Spearman rank correlations between all characteristics within the 10K human annotations while Fig.~\ref{fig:correl}b shows the p-values of these correlations. In agreement with expectations, attitudes in support of a candidate or party (``vote for,'' ``is moral'') are correlated with each other and with positive emotions (Admiration, Optimism, Joy, Pride) and are anti-correlated with their opposed attitudes (``is immoral'') and negative emotions (Anger, Embarrassment, Fear). Surprisingly, ``vote against'' is correlated with ``vote for'' possibly because there is ambiguity in whether tweets discuss voting for one candidate or against another. Positive emotions are correlated with each other as are negative emotions, in agreement with previous work \cite{alhuzali-ananiadou-2021-spanemo}, and each type of emotion is anti-correlated with its opposite. The only exception is ``amusement,'' which is correlated with negative emotions and anti-correlated with positive; this is consistent with the emotion representing sarcasm. We also find the ``economy'' concern is correlated with ``immigration,'' ``environment,'' and ``international alliances,'' while ``misinformation'' is correlated with ``international alliance.'' Finally, the ``national pride'' concern is correlated with the emotion pride.

\begin{figure}
    \centering
    \includegraphics[width=\linewidth]{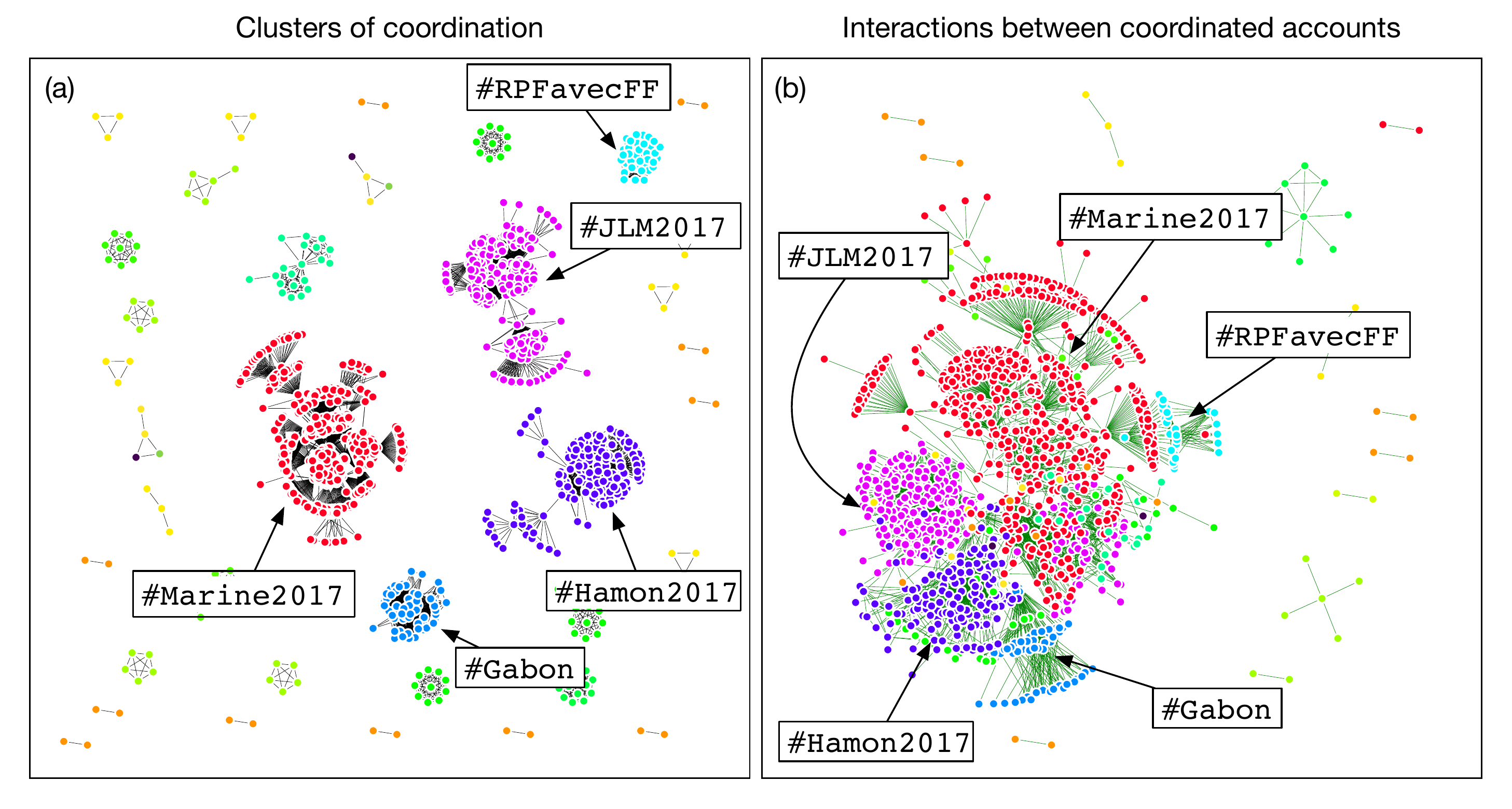} 
    \caption{Coordinated networks. (a) Nodes represent Twitter accounts and links connect accounts that share at least one original tweet with the same sequence of five or more hashtags. The most popular hashtag is listed next to the five largest connected components. (b) Retweets between coordinated accounts. Cluster colors are the same in both subfigures.
}
    \label{fig:CIAGraph}
\end{figure}
\begin{figure}[th!]
    \centering
    \includegraphics[width=\columnwidth]{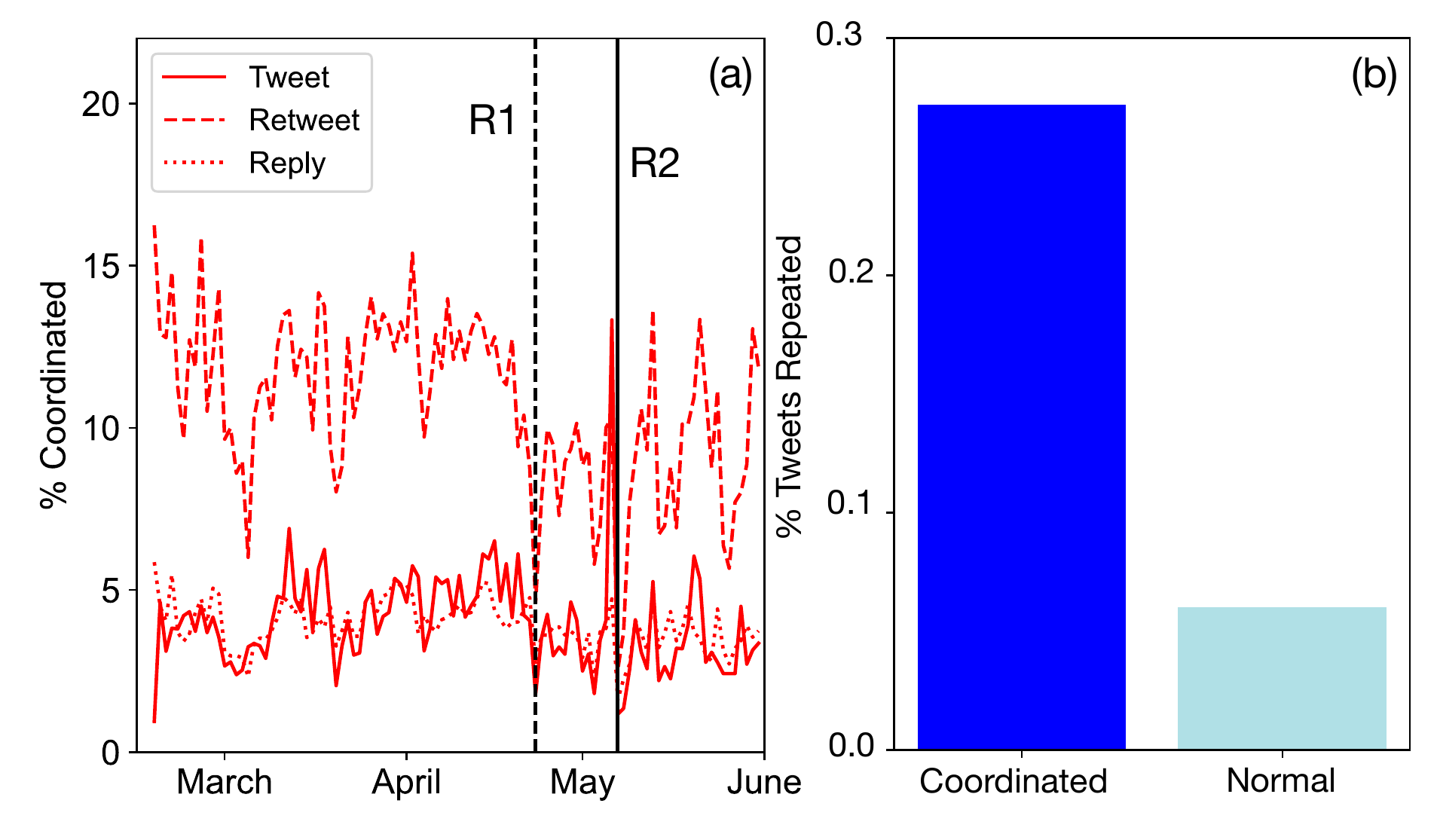} 
    \caption{Coordinated activity patterns. (a) Share of original tweets (solid line), retweets (long dashed), and replies (short dashed) that are from coordinated accounts. Another common message type, quote tweets, are not found in our dataset (which was collected by a third party), and are therefore not included in this plot. (b) Share of duplicate tweets posted by accounts.}
    \label{fig:SpamOverTime}
\end{figure}

\subsection{Coordinated Account Tactics}
We next identify networks of coordinated accounts and analyze their behavior. 

\begin{figure*}[th!]
    \centering
    \includegraphics[width=\linewidth]{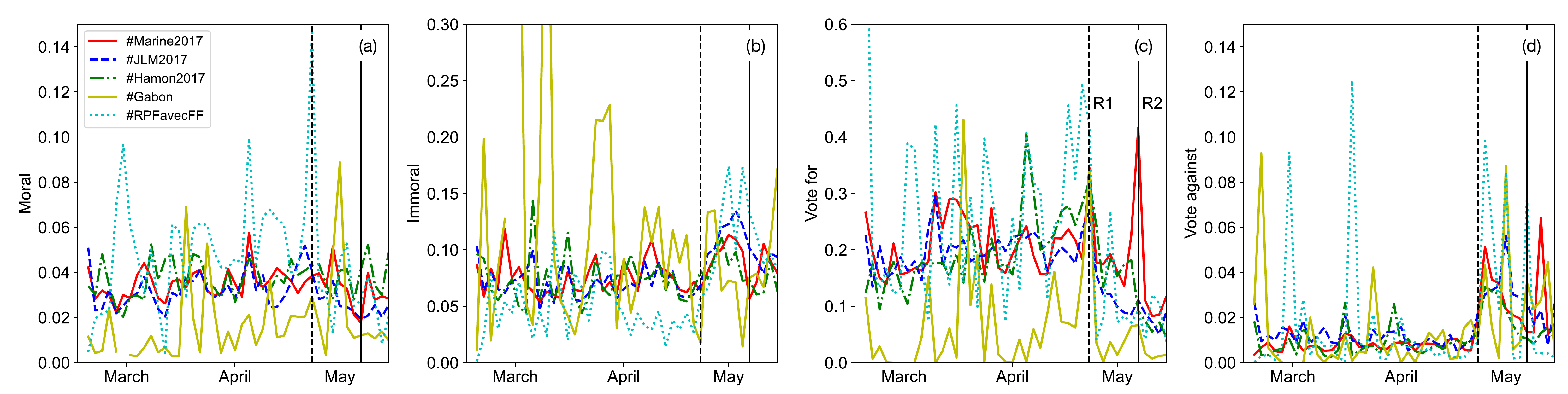}
    \caption{Attitudes over time for coordinated account clusters. Mean confidence over time for tweeting (a) a person or group is moral, (b) a person or group is immoral, (c) voting for a candidate, or (d) voting against a candidate.}
    \label{fig:ConcernsOverTime}
\end{figure*}
\subsubsection{The Network of Coordinated Inauthentic Accounts}

Fig.~\ref{fig:CIAGraph}a shows the identified network of coordinated accounts (a total of 1.6K accounts). The accounts are linked if they tweeted five or more of the same hashtags in the same order. We see several connected components, which we call coordinated account clusters. 

When we analyze the text of these coordinated accounts, we find 1.6K tweets that contained any of three hashtags often representing the \#MacronLeaks story: \#MacronLeaks, \#Bayrougate,  or \#Macrongate, while the total number of tweets in the dataset with these hashtags is 8.9K. Coordinated accounts were therefore responsible for at least 18.7\% of these conspiracy tweets despite representing just 0.28\% of all users in our dataset. In Fig.~\ref{fig:CIAGraph}b, meanwhile, we show retweets between these coordinated accounts, which shows a surprising number of interactions. In total, 10.7K retweets or 33\% of coordinated account content is retweeted by other coordinated accounts. These retweets are likely a tactic to promote content to each other's wider audiences. This tactic appears to be successful as there are 6.9K replies and 22K retweets of coordinated accounts by likely non-coordinated users.

We give an overview of coordinated account behavior in Fig.~\ref{fig:SpamOverTime}. We see in Fig.~\ref{fig:SpamOverTime}a that coordinated accounts are responsible for a disproportionate number of tweets. They represent only $0.28\%$ of all accounts yet created $\sim 5-10\%$ of tweets, replies and retweets. Just before the second round of voting, original tweets from coordinated accounts became even more prominent, possibly to promote particular candidates or to discredit Macron through \#Macronleaks. Finally, we notice a much larger proportion of coordinated account tweets were duplicates compared to normal users (Fig.~\ref{fig:SpamOverTime}b). The difference is statistically significant (Mann-Whitney U test p-value $<10^{-10}$), and our results are robust if we remove URLs or username mentions.

When we analyze individual coordinated account clusters, we notice different presidential candidates and parties are prominent. The largest cluster (927 users) used hashtags that support Le Pen (\#LePen, \#Marine2017) and often promoted conspiracies about Macron (they tweeted \#MacronLeaks 682 times, more than twenty times any other cluster). Other clusters supported Emmanuel Macron and Benoît Hamon (the three most frequent hashtags are \#Hamon2017, \#EnMarche, and \#JeVoteMacron in that order; 162 accounts) or Jean-Luc Mélenchon and La France Insoumise (the two most frequent hashtags are \#JLM2017, \#Franceincoumise in that order; 309 accounts). The latter set of coordinated accounts also promoted hashtags such as \#JulieLan\c{c}on and \#JURA, which are words related to the French 2017 legislative election on Jul 11 and 18th. Namely, Julie Lan\c{c}on was a La France Insoumise candidate in the election within Jura's 2nd constituency. Although Lan\c{c}on received only 3,323 votes in the 2017 election, at least 86\% of \#JulieLan\c{c}on tweets in our dataset (161 out of 187) were created by coordinated accounts; she was supported by 5.4\% or roughly one in twenty coordinated accounts we detected. 

We also notice a surprising cluster of 57 accounts with hashtags that include \#Gabon (the most popular hashtag), and unrelated hashtags in order of popularity \#ZDF (the German newspaper), \#10Mai2017\_A\_Geneve, and \#i, presumably to be seen in a range of Twitter conversations unrelated to Gabon. Several times the accounts mention Gabon president, such as \#BongoIsKilling (where Ali Bongo Ondimb was Gabon's president in 2017). Tweets include, ``je rêve d'un Gabon Unis sans Bongo, d'un Gabon à l'abri de la peur et du besoin \#SOSGABON...'' which translates to ``I dream of a United Gabon without Bongo, of a Gabon free from fear and need \#SOSGABON''. 

Finally, the smallest cluster is one that promotes Francis Fillon (the  three most frequent hashtags are \#RPFavecFF, \#RPF, and \#Fillon2017 in that order; 35 accounts), where \#RPF is a defunct political party, thus the account appears to want voters from a former party to vote for Fillon. Moreover, this account contains the \#Grasse hashtag, which is in reference to a shooting in the town of Grasse. The diversity of hashtags and topics within each cluster suggests that multiple, sometimes competing, influence campaigns were simultaneously active during the presidential election.

\subsubsection{Socio-Linguistic Characteristics of Coordinated Accounts}

For more insight into campaign tactics, we analyze how the mean confidences of socio-linguistic characteristics over time, where we plot attitudes over time in Fig.~\ref{fig:ConcernsOverTime}. Figure~\ref{fig:ConcernsOverTime}a--b shows that discussions of moral candidates or political parties decreases slightly between rounds one and two, but immoral claims spike just before round two. We also notice that discussions of voting for a candidate peak in round one and then decrease for clusters \#JLM2017, \#Hamon2017, and \#RPFavecFF, where the later two clusters are related to candidates who lost. Discussions about voting strongly peak in round two for \#Marine2017, suggesting a strong advocacy to vote for Marine Le Pen within that account cluster. Voting against opposition candidates across all clusters, meanwhile, peaks between rounds one and two. This also parallels analyses of emotions over time (not shown), where negative emotions peak between round one and two. This agrees with analysis of 10K human annotated data, where we find anger, fear, and negative-other correlates with voting against a candidate (Spearman rank correlations$=0.06,~0.03,~0.08$, p-values$\le0.001$).

While model confidences are used throughout the paper, we can also binarizing labels. For example, a tweet with confidence 0.8 that it contains the love/admiration emotion is then given the label 'love/admiration'. This results in virtually identical results. Namely, while the Spearman correlation between confidences and binarized labels across all 5M users aggregated at the daily level vary for each socio-linguistic characteristic, the median correlation is high at 0.85.
\begin{figure*}
    \centering
    \includegraphics[width=\linewidth]{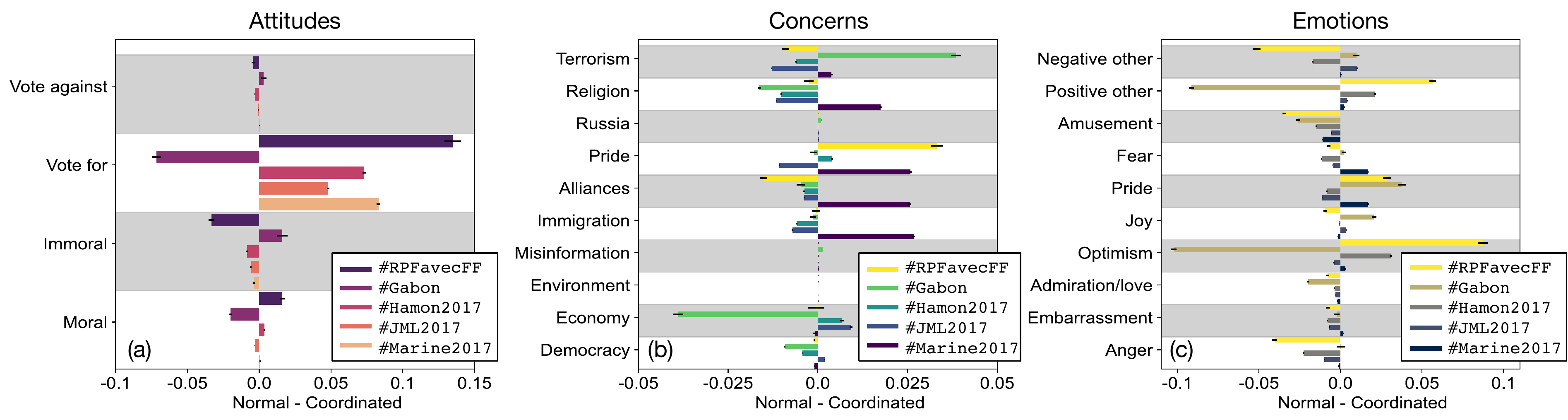}
    \caption{Socio-linguistic characteristics used by coordinated information campaigns. (a) Attitude, (b) concerns, and (c) emotions for each of the five largest coordinated account clusters (sorted from top to bottom: \#RPFavecFF, \#Gabon, \#Hamon2017, \#JML2017, and \#Marine2017). The x-axis shows the difference between the mean tweet confidence of each cluster compared to non-coordinated campaign tweets. Positive values indicate coordinated account tweets whose socio-linguistic characteristic confidences are higher than non-coordinated accounts, and negative values indicate confidences that are lower. Black lines represent standard errors. 
}
    \label{fig:CIAdiff}
\end{figure*}

\begin{figure*}[tbh!]
    \centering
    \includegraphics[width=\linewidth]{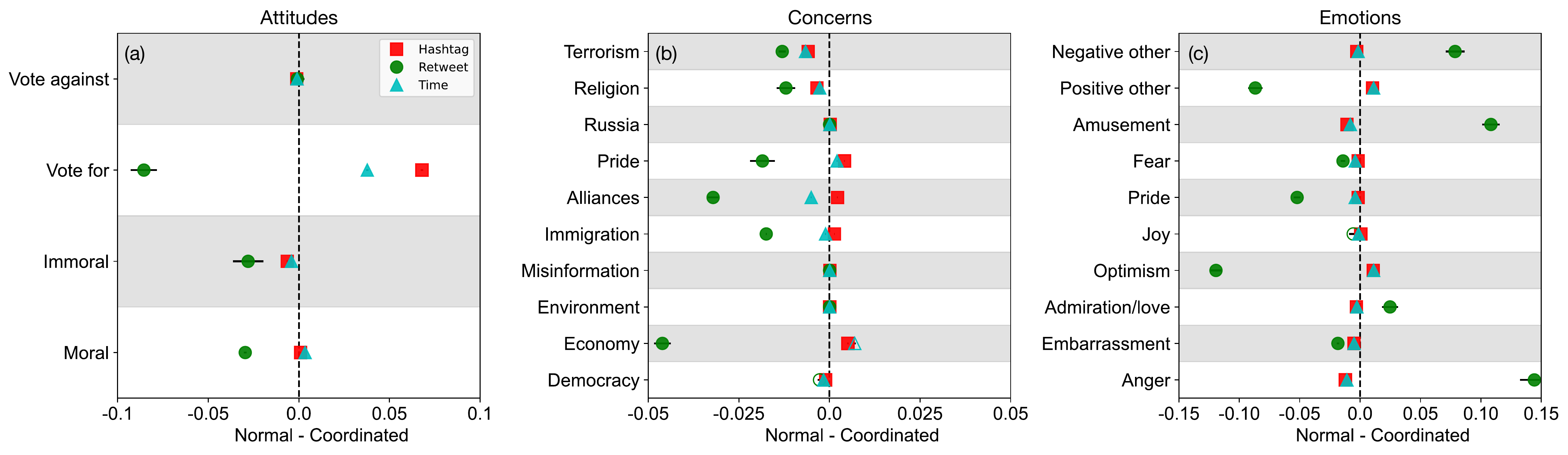}
    \caption{Socio-linguistic characteristic confidence differences between normal and coordinated networks. Coordination is defined as sharing unique sequences of hashtags in a tweet (``hashtag'' - 1.6K users), similarities of retweets (``retweet'' - 24 users), or similarities of tweet times (``time'' - 404 users). (a) Attitudes, (b) concerns, and (c) emotions. Black lines represent standard errors. Open markers represent values not statistically significantly different from 0 (Mann-Whitney U test p-values $>0.05$).}
    \label{fig:CIArobust}
\end{figure*}

We next analyze time averaged socio-linguistic characteristics within coordinated account clusters in Fig.~\ref{fig:CIAdiff}. This figure takes the difference in the mean tweet confidence between accounts withing a coordinated network or cluster and all ordinary (non-coordinated) accounts. All results are significant based on the Mann-Whitney U test (p-values $<0.05$) except: the attitude ``vote against'' for \#RPFavecFF, the concern ``alliances'' for \#Gabon, and the emotions ``negative-other'' for \#RPFavecFF, and ``anger'' and ``embarrassment'' for \#JML2017. There are many similarities across coordinated networks. First in Fig.~\ref{fig:CIAdiff}a, coordinated accounts tweet more about the voting for candidates (vote for attitude). To put these values in perspective, if we binarize labels for each tweet, we find that 35\% of all coordinated account tweets promote a candidate or party in contrast to just 8.2\% among non-coordinated users. In Fig.~\ref{fig:CIAdiff}b, we find that larger coordinated account clusters have lower religion, alliances, and immigration confidences, with the exception of the \#Marine2017 cluster. Finally, in Fig.~\ref{fig:CIAdiff}c, coordinated accounts tend to have lower amusement, embarrassment, and admiration/love confidences. 

Key differences, however, abound. Most notably the \#Gabon cluster's attitudes have lower voting or positive moral stances confidences but a higher immoral stance. Meanwhile, their terrorism concern confidences are higher, and economic concern confidences are lower than non-coordinated accounts. Finally, their tweets are very negative with low optimism or positive emotions. This reflects their typically off-topic and admonishing tweets about Gabon's president. The \#Marine2017 cluster, meanwhile is unusual by having higher religion, national price, alliance, and immigration confidences than non-coordinated users (and most coordinated clusters). The \#Marine2017 cluster therefore appears to be diving deep into divisive issues, perhaps to separate Marine from other candidates or perhaps to create wedge issues that divide the electorate.

There are a number of coordination metrics \cite{Pacheco2021}, therefore, to check the robustness of our results, we also determined coordination based on similarities of retweets (24 accounts, no accounts overlap with hashtag-based coordinated accounts), and the timing of tweets (404 accounts, 108 overlapping with hashtag-based coordinated accounts). The results are summarized in Fig.~\ref{fig:CIArobust}, where we take the difference in the mean confidences between coordinated and non-coordinated accounts. We find consistent behavior between hashtag and tweet time-based coordinated accounts, where about 27\% of the tweet time-based set of coordinated accounts are also in the hashtag-based set of coordinated accounts. Retweet-based accounts show distinct behavior both because of the small number of (possibly non-representative) accounts and because the accounts may utilize a different set of manipulation tactics.

\begin{figure}
    \centering
    \includegraphics[width=\columnwidth]{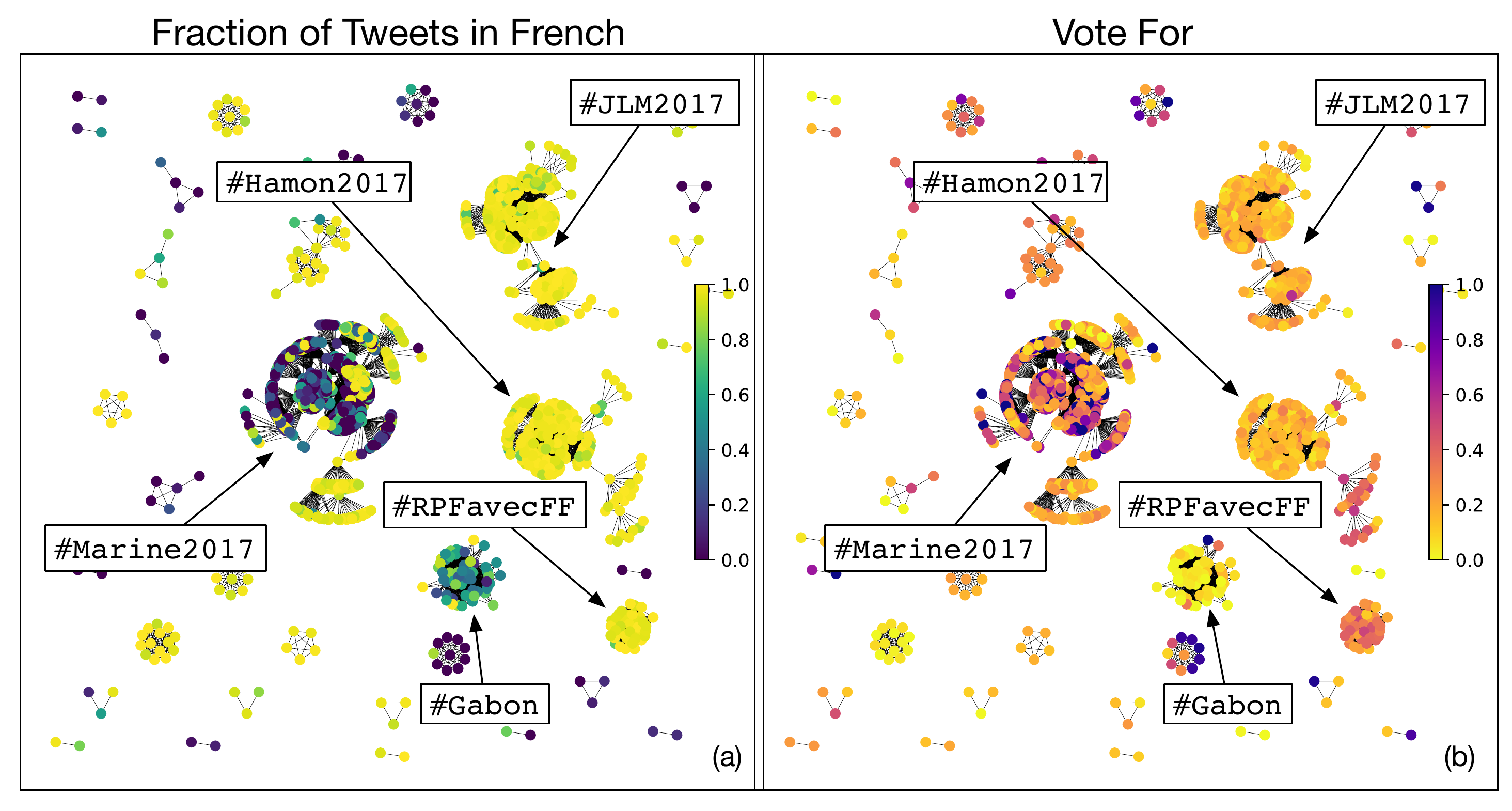}
    \caption{Socio-linguistic characteristics within information campaigns. Mean tweet confidence per user for (a) fraction of tweets in French and (b) vote for attitude.}
    \label{fig:CIAnet}
\end{figure}

Not only can we capture cluster-level behavior, our analysis can also reveal differences in individual coordinated accounts, which we show in Fig.~\ref{fig:CIAnet}. 
Several findings are apparent in Fig.~\ref{fig:CIAnet}a. First, the \#Marine2017 cluster stands out for having surprisingly few tweets in French (whose language is indicated by the Twitter API), with only 36\% of tweets in French on average for each account while 48\% are in English. This agrees with previous finding that a majority of tweets in the \#MacronLeaks campaign were in English \cite{vilmer2021fighting}. There are also many non-French accounts in the \#Gabon cluster, although there is greater uniformity. Next, we demonstrate the diversity of socio-linguistic characteristics across accounts with a case study in Fig.~\ref{fig:CIAnet}b, which shows the mean vote for attitude confidence across all tweets for each coordinated account. The confidence is especially high for the \#Marine2017 cluster although there is a wide variance. In contrast, the \#Gabon cluster has very low vote for attitude confidence across all accounts. 

\subsection{Discussion}
The results demonstrate key differences in socio-linguistic characteristics of coordinated accounts and non-coordinated accounts, which provides a nuanced understanding of coordinated accounts. First, these results show that coordinated accounts retweet a disproportionate amount, especially retweeting each other to amplify their messages, and repeat their tweets more often than ordinary users in order to amplify exposure through repeated messaging. This is a tactic useful to increase online attention \cite{cox2002beyond}. Next, the socio-linguistic characteristics demonstrate that coordinated accounts attempt to push a ``vote for candidate'' message far more than non-coordinated accounts, especially before elections, possibly to guide potential voters to a very specific candidates. Interestingly, we also found negative emotions increased between election rounds. Negative campaigns can be effective if done correctly \cite{Fridkin2004}, and attacks against candidates could be more believable \cite{Fessler2014}. Moreover, coordinated accounts selectively push particular election concerns (most notably, the \#Marine2017 coordinated cluster discussed national pride, alliances, and immigration). The results also show coordinated accounts promoting small elections, such as the candidate Julie Lan\c{c}on. The outlier \#Gabon cluster meanwhile does not seem to advocate for a particular candidate but instead mentions prominent Twitter accounts  (every single \#Gabon account tweet mentions a Twitter user), including French politicians. This may be a tactic for these tweets to appear in Twitter searches for the politician’s Twitter handle (an especially likely scenario during an election). This will allow a wider international audience to see these messages.

Our work highlights a number of wider implications. Namely, coordinated accounts have remarkable diversity the agendas, concerns, and emotions they share, even within closely aligned clusters. There is no exemplar markers of influence campaigns, presumably because these coordinated accounts attract a different audience. Related to this, the type of tweets vary over time; for elections this can be to promote (vote for) or attack (vote against) parties and candidates, yet coordinated account clusters often retweet each other (in agreement with a previous paper \cite{Wang2023}). This may be a tactic boost each other's messages.

\section{Conclusions}
Our analysis of a large body of tweets related to the 2017 French election reveals psycho-social dynamics of coordinated accounts. While the coordinated accounts we identified were only 0.28\% of all users, they comprised of 5-10\% of all retweets and at least 18.7\% of \#MacronLeaks tweets, an information campaign led by Russia. Consistent with this, we also find coordinated account activity spiked just before round two (when the \#MacronLeaks story first appeared). Coordinated accounts appear to have employed a range of tactics, such as repeating their messages, sharing positive content (``vote for'' rather than ``vote against''), sharing more positive emotions, and focusing on some voter concerns, such as national pride and the economy. That being said, we also notice a degree of diversity in coordinated account clusters, possibly because these clusters are tailored to different audiences. Overall, the results point to coordination accounts being used for social manipulation and we uncover potential tactics towards that purpose.

While our methods have given new insights into coordinated accounts, they have a number of limitations that motivate future work. Namely, we find the socio-linguistic characteristic models are imperfect. This is a limitation, which should be improved in the future. Part of this limitation is due to data imbalance, therefore more data, especially for low-support classes is critical. Next, the data is a biased sample \cite{morstatter2013sample}, which limits the generalizability of our findings. A more representative sample, especially of recent elections is needed to validate these findings. In addition, the degree to which these results generalize outside France or outside of elections needs to be studied. Finally, the coordinated account metrics are imperfect because we do not have ground truth labels. While different coordination metrics show the robustness of our results, these metrics are not perfect indicators of coordination. Future work is therefore needed to train models on ground truth data. It will be especially useful to detect the type of coordination (retweeting the same content, versus repeating tweets, versus sharing tweets in synchronized times, etc.) which may be a factor in how coordinated accounts behave.

\subsection*{Broader perspective, ethics and competing interests}

All data is public and collected following Twitter's terms of service, with the study considered exempt by the authors' IRB. To minimize risk to users, all identifiable information was removed and analysis was performed on aggregated data. We therefore believe the negative outcomes of the use of these data are minimal. 

Our analysis of these data will have broad positive impact in understanding tactics of information campaigns. Researchers can use these findings to potentially better identify information campaigns in the future and reduce the harm they continue to pose. There is a chance that knowledge of these tactics could entice bad actors to change or hide their behavior, but we believe the benefit of transparency outweighs this risk. While these tweets are related to the 2017 French election, we expect our findings to generalize to other political scenarios.


\end{document}